# Superconductivity and Superconductor-Insulator Transition in Single Crystal Sb$_2$Te$_3$ Nanoflakes

Wei-Han Tsai,[†] Chia-Hua Chien,[†,⧧] Ping Chung Lee,[†] Min-Nan Ou,[†] Sergey R. Harutyunyan,[*,†,‡] Yang-Yuan Chen[†,#].

[†]Institute of Physics, Academia Sinica, Nankang, Taiwan

[⧧]Department of Engineering and System Science, National Tsing Hua University, Hsinchu, Taiwan

[#]Graduate Institute of Applied Physics, National Chengchi University, Taipei, Taiwan

[‡]Institute for Physical Research, NASRA, Ashtarak, Armenia

The discovery of the topological insulators offers a great opportunity to search for a superconducting analog, a time–reversal–invariant topological superconductor. We report on transport properties of the topological insulator single crystal Sb$_2$Te$_3$ nanoflakes with thickness about 7–50 nm. A steep drop of resistance is appeared near 3 K in the ultrathin Sb$_2$Te$_3$ nanoflakes, manifesting a superconducting transition. The results show that the existence of certain optimum degree of disorder is a necessary condition for emergence of superconductivity. The magnetic-field–induced superconductor–insulator transition of disordered 2D superconductor system is observed in the nanoflakes. Temperature dependence of magnetoresistance shows a consecutive transformation of weak anti-localization cusp into the superconducting transition at low field when $B < B_C$.

After the discovery of the topological insulators, TIs, it was realized that there are strong correlations between superconductivity and topology nature. Topological insulators (TIs) are a new class of quantum matters, in which gapless, metallic surface states (SS) coexist with the bulk band-gap region.[1–3] The surface states originate due to the strong spin-orbit (SO) interaction, composed of spin-momentum locked helical massless Dirac cones protected by time reversal (TR) symmetry. A remarkable consequence of this is the prevention of the surface state electrons from back scattering caused by imperfections of the lattice. A direct analogy exists between superconductors and insulators because the Bogoliubov–de Gennes (BdG) Hamiltonian for the quasiparticle excitations in a superconductor is analogous to the Hamiltonian of a band insulator, with the superconducting gap corresponding to the band gap of the insulator. This leads to possibilities in realization of novel superconductivity.[4–6] Such topological superconductor is associated with quasiparticle excitations which are Majorana fermions and obey non-Abelian exchange statistics that opens up new approaches for future quantum devices and computations.[7–10] Another important aspect is the possibility to observe superconductivity in disordered 2D systems. This subject has become even more relevant with the realization that high-transition-temperature (high-Tc) superconductors are intrinsically disordered.[11]

By now, a limited number of materials, which can be attributed to the topological superconductors such as $Cu_xBi_2Se_3$,[12,13] $Bi_2Te_3$ and $Sb_2Te_3$ under high pressure,[14,15] $In_xSn_{1-x}Te$,[16] $Cu_x(PbSe)_5(Bi_2Se_3)_6$,[17] $Sr_xBi_2Se_3$,[18] $Tl_5Te_3$,[19] and $Sb_2Te_3$,[20] grown under pressure with excess amount of Te, have been discovered and studied. The superconductivity obtained in our $Sb_2Te_3$ nanoflakes is ascribed to the variation of thickness of nanoflakes. This opens up new ways to understand the interrelation between the topological surface states and superconductivity.

The $Sb_2Te_3$ single crystal compound studied in the present work does not belong to superconductors and is a robust topological insulator.[1, 2, 21] The compound possesses rhombohedral crystal structure with the space group $D_{3d}^5(R\bar{3}m)$. The Sb$_2$Te$_3$ single crystal is naturally *p*–doped (the Sb vacancies and Sb$_{Te}$ anti–site defects are responsible for the hole generation), and the Dirac point is located above the Fermi energy $E_F$.

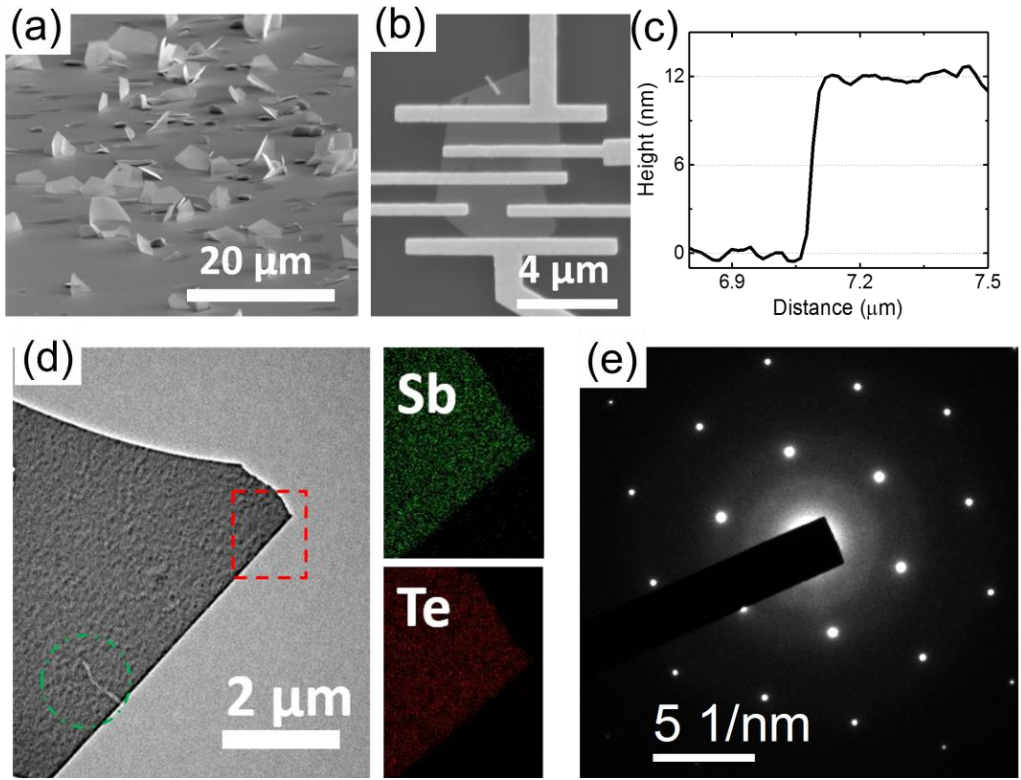

**Figure 1.** (a) Grown nanoflakes on $SiO_2$/Si substrate. (b) Sample NF12 with deposited electrodes. (c) The height profile of the NF12 nanoflake obtained by AFM. (d) TEM image and EDX mapping picture for one of the samples, the circle area shows the existence of crack on the nanoflake. (e) TEM SAED pattern of the sample. Notation: The number after NF represents the thickness of nanoflake, for example NF12 means the thickness of the nanoflake is 12 nm.

In this research, the vapor transport method is employed to synthesize Sb$_2$Te$_3$ nanoflakes of thicknesses, *d,* about 7–50 nm, measured with the accuracy ± 0.5 nm. The SEM image of single crystalline Sb$_2$Te$_3$ nanoflakes of various thicknesses grown on SiO$_2$/Si substrate is demonstrated in **Figure 1a**. The TEM analysis and the EDX mapping picture of one of the grown samples characterize the nanoflakes as single crystal with uniform spatial distribution of Sb and Te (**Figure 1d,e**). The atomic ratio of Sb/Te corresponds to (40±1)/(60±1). Our experiments demonstrate an onset of superconductivity in the ultrathin single crystal Sb$_2$Te$_3$ nanoflakes. The 2D superconductivity revealed in these ultrathin samples results in the observed magnetic-field-tuned superconductor-insulator transition, SIT.

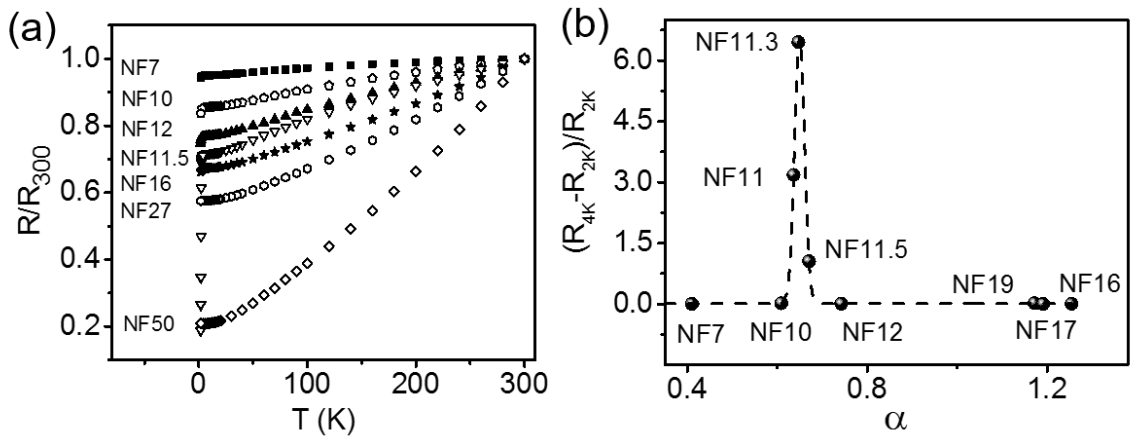

**Figure 2**. (a) The normalized resistances of samples with various thickness in the region of 1.8−300 K. (b) The relative change of resistance at superconducting transition depending on parameter α for samples with various thickness. Dash line is guide for eyes.

The curves of normalized resistance $R(T)/R_{300}$ for several nanoflakes of different thicknesses are shown in **Figure 2a**, where $R_{300}$ is the resistance at 300 K. Decreasing slope of $R(T)/R_{300}$ curves with thickness is a consequence of increasing disorder in the samples. The resistances of the nanoflakes thinner than 27 nm (the number after NF represents the thickness of nanoflake), drop steeply near 3 K, represent an onset of the superconducting transition. The magnitude of the relative drop of the resistance at the superconducting transition depends on degree of disorder of the nanoflakes. However, the degree of disorder does not correlate unambiguously with the thickness and resistivity of nanoflakes (at least in the limit of several nanometers correction). This may result from unintentional damages (cracks, holes, partly torn-off surface) accumulated during the preparation of the samples. Thus, we consider the normalized resistance as the most reliable parameter to compare different nanoflakes. For the nanoflakes with superconducting transition, the $R(T)/R_{300}$ curves behave as a power law function $R \sim T^{\alpha}$ in the strong electron-phonon interaction region (50–300 K), where the α parameter is around 0.4–1.25. For the samples with highest content of a superconducting phase, having larger relative change of resistance $(R_{4\,K} - R_{2\,K})\,/\,R_{2\,K}$ at superconducting transition (**Figure 2b**), the parameter α varies in the limits of $0.65 \pm 0.03$, which implies existence of an optimum value of the degree of disorder.

Magnetoresistances for some of the samples with different thicknesses, at 2 K are presented in **Figure 3**. There is a noticeable difference in MR depending on a thickness of the nanoflakes. MR of thick samples NF50 and NF27 contains considerable Lorentz type contribution (proportional to $(\mu B)^2$, where μ is mobility of charge carriers) that is the response of the bulk states. The NF27 nanoflake already shows a weak antilocalization (WAL) cusp on the MR curve. In the thinner samples, the Lorentz type MR practically does not exist. Instead, nearly constant MR in high magnetic fields is observed and the cusp is "enormously" large, assuring almost total variation of the magnetoresistance. Such behavior suggests that near 2 K, in low fields, almost the whole contribution to the conductivity of these ultrathin nanoflakes comes from the surface states and from the superconductivity through the formation of Cooper pairs.

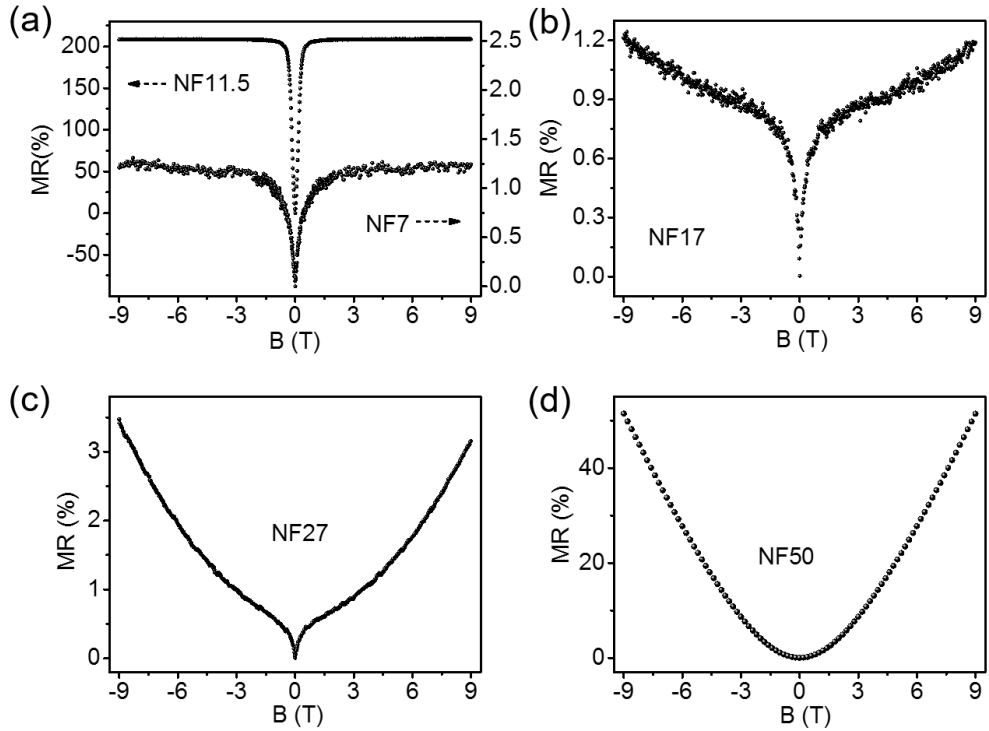

**Figure 3.** (a) Magnetoresistance of NF7 sample and NF11.5 sample. (b) Magnetoresistance of NF17 sample. (c) Magnetoresistance of NF27 sample. (d) Magnetoresistance of NF50 sample.

In high magnetic fields, the charge carriers become localized in the ultrathin nanoflakes due to suppression of the superconductivity and broken TR symmetry. The evolution of longitudinal resistance $R_{xx}(B)$ (typical for all nanoflakes, experiencing superconducting transition) with temperature in transverse magnetic fields for the samples NF7 and NF11.5 is demonstrated in **Figure 4a,b**. The $R_{xx}(B)$ curves are crossed at some definite magnetic field region $B_C$ and eventually become arranged in the inverse order at fields $B > B_C$, that is, the magnetoresistance at the given magnetic field decreases when the temperature increases. This is the behavior of the magnetic-field-induced superconductor-insulator transition in the limit of two dimensions and zero temperature, observed in thin films.[22–24] The transition is a quantum phase transition, dominated and controlled by quantum fluctuations. Tuning of SIT is achieved by applying a transverse magnetic field as well as by varying film thickness. Magnetic field transforms the superconducting state of a disordered film at low fields through a metallic one at the critical field $B = B_C$ to an insulating state at $B > B_C$.

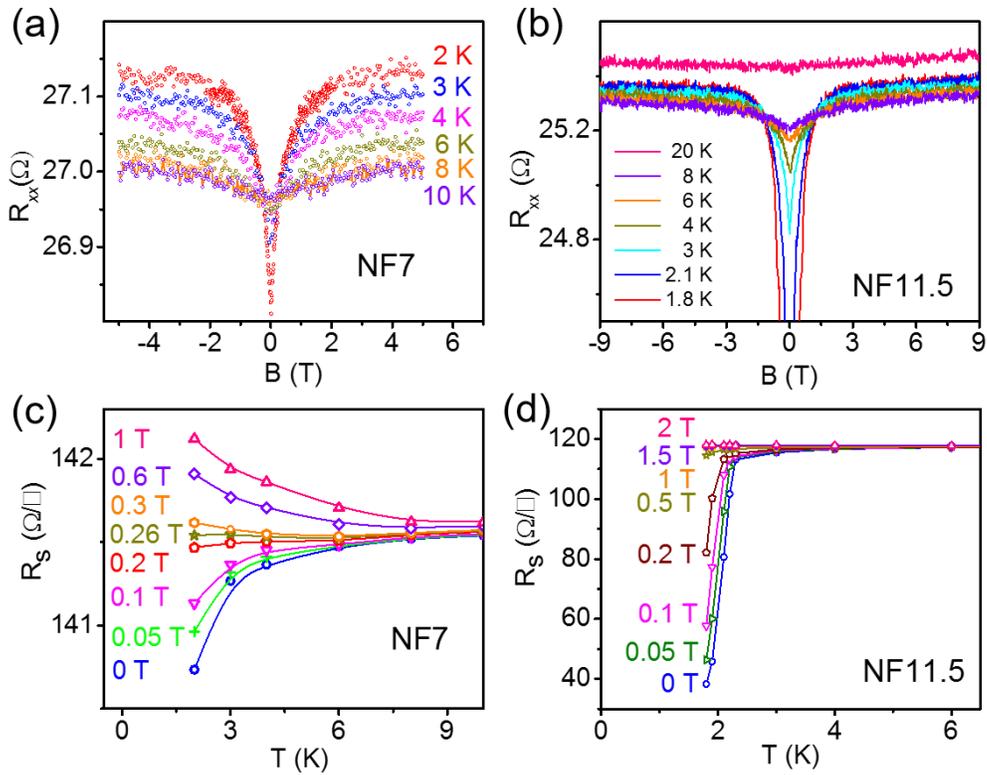

**Figure 4.** (a) $R_{xx}(B)$ curves of the NF7 sample, at different temperatures. (b) $R_{xx}(B)$ curves of the NF11.5 sample, at different temperatures. (c) $R_S$ vs temperature of the NF7 sample. (d) $R_S$ vs temperature of the NF11.5 sample.

The set of isomagnetic curves for the samples NF7 and NF11.5 at different transverse magnetic fields, shown in **Figure 4c,d**, can be divided into two groups by the sign of $dR/dT$. The positive (negative) sign of the value corresponds to the superconducting (insulating) behavior. The boundary isomagnetic curve $R_C(T)$ between superconductor and insulator phases corresponds to the boundary metallic state at T = 0 K and indicates the critical field value $B_C$. We plot the surface resistance $R_S$ against the scaling variable $|B - B_c|/T^{1/zv}$, where $v$ is the correlation length exponent and z is the dynamical-scaling exponent. The product of critical exponents varies in the region $zv = 0.75 \pm 0.05$ for different samples and is obtained through the best collapse between measured data (**Figure 5a,b**). At a finite temperature, the size of quantum fluctuations is restricted by the dephasing length $L_\varphi \propto T^{-1/z}$, where $z$ is expected to be equal to 1 for SIT. Assuming that $z = 1$, we get $v = 0.75 \pm 0.05$, which is close to the values obtained for amorphous Bi films. The value of exponent is inconsistent with the scaling theory, which predicts $v > 1$. Previous studies of SIT have revealed that when the product of exponents is determined by magnetic field rather than the thickness as tuning parameter, $zv = 0.7 \pm 0.2$.

The value of the critical resistance $R_C$ of the nanoflakes varies in the region between 100 Ω/□ and 200 Ω/□, depending on samples and deviates from the universal value $R_Q = h/4e^2 \approx 6.5$ kΩ/□ predicted by the scaling theory, and may reflect the existence of an additional conducting channel. The correction from the quantum interference for temperature dependent magnitude of conductivity in 2D is given as $\Delta\sigma \sim \pm p\{ln(T/T_L)\}$, where $T_L$ is the characteristic temperature, $+p$ corresponds to weak localization (WL), $-p$ to WAL and $p$ is a constant which depends on dephasing mechanism.[25] **Figure 5c,d** shows that the temperature dependence of the conductivity correction in high magnetic fields, $B > B_C$, approaches to ~ $ln(T)$ i.e. demonstrates WL effect. Weakening of the magnetic field leads to increasing contribution of the WAL effect. In low fields, where the superconductivity becomes well developed, the dependence departs from pure WAL and deviates from ~ - $ln(T)$.

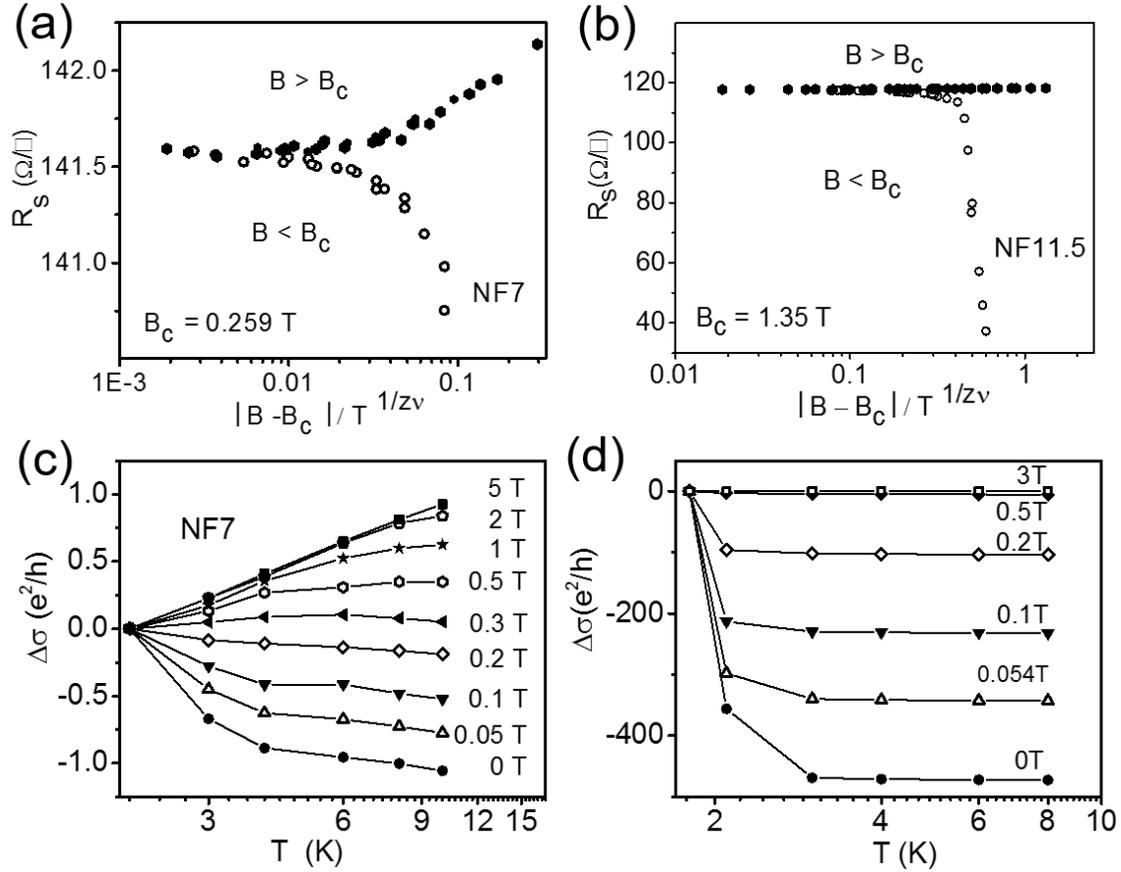

**Figure 5**. (a) Scaling of $R_S$ vs $|B - B_C|/T^{1/zv}$ for sample NF7. (b) Scaling of $R_S$ vs $|B - B_C|/T^{1/zv}$ for sample NF11.5. (c) Conductivity correction for sample NF7. (d) Conductivity correction for sample NF11.5.

The angular dependences of the resistance $R(\Theta)$ in magnetic field obtained for one of the samples demonstrate absence of any response to the parallel magnetic field and confirm 2D nature of superconductivity in the nanoflake (**Figure 6**).

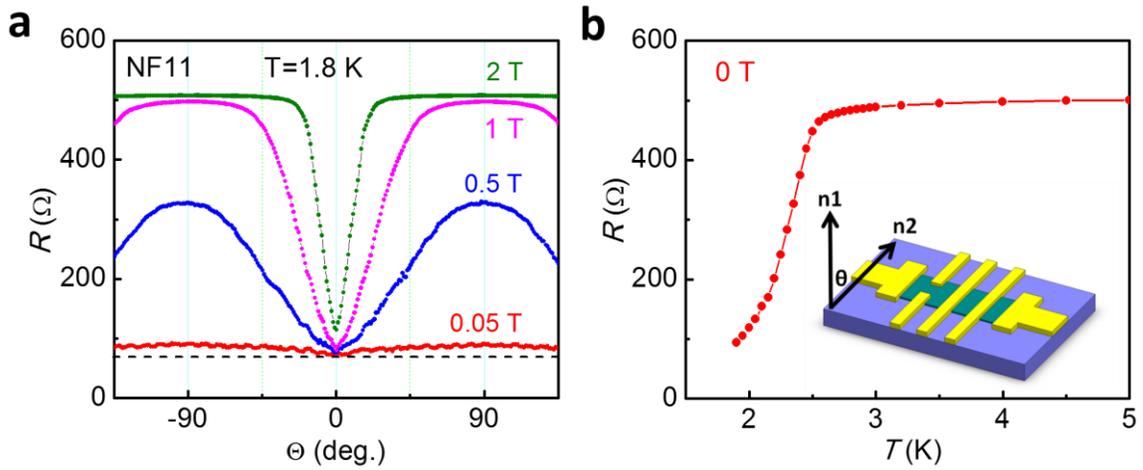

**Figure 6**. (a) $R(\Theta)$ dependence of NF11 sample at 1.8K in different constant magnetic fields. Dashed line shows the resistance value at 0 T. (b) $R(T)$ at superconducting transition region; the inset shows the nanoflake on substrate relative to the rotation direction of the magnetic field induction vector ***B***.

The bulk states in a topological insulator thin film can exhibit weak localization in confined dimension when splitting of 3D bulk bands into 2D subbands occurs.[25, 26] In our samples, WL is masked by strong contribution into the conductivity from the surface states and superconductivity. The weak localization appears only in high magnetic fields, when TR symmetry is broken and the superconductivity is suppressed. WL is a consequence and an indicator of the disorder. The obtained $R(T)$ curves (**Figure 2**) show not only the existence of some optimum value of disorder (localization of the bulk carriers), but also some necessary degree of disorder for superconducting transition to emerge. In other words, with other conditions being equal, there is a boundary thickness (or disorder) at which the bulk states become partly localized and the Cooper pairs start to develop. This condition implies an enhanced contribution from the surface states in comparison with thicker samples. In the thinnest samples, we observe the weakening of superconductivity. The reason of such a behavior may be the excess disorder and the weakened impact of the surface states caused by a deep hybridization between top and bottom surface states. When the penetration length $\xi = hv_f/2\pi M$, where $M$ is the bulk band gap and $v_f$ is the Fermi-velocity, becomes comparable to film thickness, then a gap $\Delta$ opens up at the $\Gamma$ point and the dispersion of SS changes to Dirac-hyperbolas, which results in crossover to WL.[6, 25]

Thus, we can suppose that the combined effect of the localized bulk states and the surface states is a necessary condition for the origin of 2D superconductivity in $Sb_2Te_3$ topological insulator. Despite the evident deviations of $R_C$ and $\nu$ parameters from those predicted by SIT theory, the main characteristics of the quantum SIT i.e. the $B_C$ critical point and scaling behavior are preserved for the $Sb_2Te_3$ nanoflakes. Temperature dependent isomagnetic curves obtained for conductivity correction $\delta\sigma_L(T)$ clearly show gradual transformation of WAL effect into superconducting transition. Moreover, successive fitting of the magnetoconductivity curves by HLN equation[27] far from SC transition confirms the existence of WAL in the $Sb_2Te_3$ nanoflakes. The existence of WL–WAL crossover and transformation of WAL into SC transition in $Sb_2Te_3$ is unique and opens new ways for understanding the interrelation between the topological surface states and superconductivity.

In conclusion, we report on transport properties of the topological insulator $Sb_2Te_3$ nanoflakes of different thickness. 2D superconductivity and magnetic-field-tuned superconductor-insulator transition were observed in the ultrathin samples. The consecutive transformation of the superconducting transition to pure WAL effect with increasing temperature in the ultrathin nanoflakes was observed. It is supposed that the mutual effect of bulk carrier's localization and the surface states are the necessary condition for the origin of 2D superconductivity in the $Sb_2Te_3$ topological insulator.

We would like to thank Dr. Paul C.W. Chu, University of Houston, Dr. Ting-Kuo Lee, Dr. Maw-Kuen Wu and Dr. Yip Sungkit, Academia Sinica for their valuable suggestions and discussion, and the assistance of Core Facilities for Nanoscience and Nanotechnology at Academia Sinica. This work was supported by the Ministry of Science and Technology under Grant No. NSC 103-2112-M-001-021-MY3.

*sergeyhar56@gmail.com